\documentclass[aps,prd,nofootinbib,twocolumn,floatfix,superscriptaddress,eqsecnum]{revtex4-1}
\usepackage{amsmath,amssymb,amsthm,bm,bbm}
\usepackage{graphicx,psfrag,subfigure,rotating}

\usepackage{mycom}

\begin{document}
\title{Lattice Index Theorem and Fractional Topological Charge}
\author{R. H\"ollwieser}
\affiliation{Atomic Institute, Vienna University of Technology,
Wiedner Hauptstr.\ 8-10, A-1040 Vienna, Austria}

\author{M. Faber}
\affiliation{Atomic Institute, Vienna University of Technology,
Wiedner Hauptstr.\ 8-10, A-1040 Vienna, Austria}

\author{U.M. Heller}
\affiliation{American Physical Society, One Research Road,
Ridge, NY 11961, USA}

\date{\today}
\begin{abstract}
We study topological properties of classical spherical center vortices with the low-lying eigenmodes of the Dirac operator in the fundamental and adjoint representations using both the overlap and asqtad staggered fermion formulations. In particular we address the puzzle raised in a  previous work of our group [Phys.\ Rev.\ D 77, 14515 (2008)], where we found a violation of the lattice index theorem with the overlap Dirac operator in the fundamental representation even for ``admissible'' gauge fields. Here we confirm the discrepancy between the topological charge and the index of the Dirac operator also for asqtad staggered fermions and the adjoint representation. Numerically, the discrepancy equals the sum of the winding numbers of the spheres when they are regarded as maps $\mathbbm R^3 \cup \{\infty\} \to SU(2)$. 
Furthermore we find some evidence for fractional topological charge during cooling the spherical center vortex on a $40^3 \times 2$ lattice. The object with topological charge $Q=1/2$ we identify as a Dirac monopole with a gauge field fading away at large distances. Therefore even for periodic boundary conditions it does not need an antimonopole.
\end{abstract}
\pacs{11.15.Ha, 12.38.Aw}
\keywords{Lattice Gauge Field Theories, Topological Charge, Vortices, Monopoles}
\maketitle

\section{Introduction}

Topological models have been introduced to explain quark confinement in
non-abelian gauge theories. There are good reasons to believe that a force
strong enough to confine quarks must also break chiral symmetry
spontaneously~\cite{Casher:1979vw}. The center vortex
model~\cite{'tHooft:1977hy,Vinciarelli:1978kp,Yoneya:1978dt,Cornwall:1979hz2,Mack:1978rq,Nielsen:1979xu}
seems to be the most promising candidate to explain both phenomena. Center
vortices, quantized magnetic flux lines, compress the gluonic flux into
tubes and cause a linearly rising potential at large separations. Numerical
simulations have indicated that vortices could also account for phenomena
related to chiral symmetry, such as causing topological charge fluctuations
and spontaneous chiral symmetry breaking
(SCSB)~\cite{Reinhardt:2000ck2,deForcrand:1999ms,Alexandrou:1999vx,Engelhardt:2002qs}.
These nonperturbative features of the QCD vacuum are intimately linked to
the properties of the low-lying spectrum of the Dirac operator.  In
Ref.~\cite{Hollwieser:2008tq2} we found strong correlations between the
center vortices and the density distribution of low-lying Dirac eigenmodes.
We showed that even center-projected lattice configurations give rise to a
spectral density of near-zero modes, which by the Banks-Casher
relation~\cite{Banks:1979yr} is proportional to the chiral condensate, the
order parameter for SCSB.  Another relation between fermions and topology
is given by the Atiyah-Singer index theorem
\cite{Atiyah:1971rm,Schwarz:1977az,Brown:1977bj,Narayanan:1994gw}. It
states that the topological charge of a gauge field configuration equals
the index of the Dirac operator in this gauge field background. In
Ref.~\cite{Jordan:2007ff} we investigated the lattice index theorem and the
localization of the zeromodes for thick classical center vortices. For
nonorientable spherical vortices, the index of the overlap Dirac operator
turned out to differ from the topological charge. This may be related to
the fact that even in Landau gauge some links of this configuration are
close to the nontrivial center elements.

In this paper, we work with thick spherical vortices in SU(2) lattice gauge
theory and extend our analysis of the topological charge and the lattice
index theorem. In Sec.~\ref{sec:topcharge} we re-introduce the construction
of a spherical vortex and report on our calculations with the overlap Dirac
operator. After analyzing the response of (improved) staggered fermions to
the topological background of the gauge field we investigate in
Sec.~\ref{sec:topstagg} the localization of asqtad staggered zeromodes with
respect to the position of the thick vortices and find the same discrepancy
for the topological charge determined by different methods.  We then
examine the index theorem for adjoint fermions in Sec.~\ref{sec:adjoindex}
and discuss the role of adjoint fermions with respect to fractional
topological charge in Sec.~\ref{sec:fractadjoint}.  We find an interesting
configuration giving topological charge one half which we analyze in
Sec.~\ref{sec:fracttconfig} in more detail. We end with our conclusions.

\section{Spherical Center Vortices and Dirac zero modes}
\label{sec:topcharge}

The nonorientable spherical vortex of radius $R$ and thickness $\Delta$ is
constructed with the following links:
\begin{gather}
U_\mu(x_\nu) = \begin{cases}
 \exp\left(\mathrm i\alpha(|\vec r-\vec r_0|)\vec n\cdot\vec\sigma\right)&t=1,\mu=4\\
 \mathbbm{1}&\mathrm{elsewhere}\end{cases}\\ \qquad\mbox{with}\quad
 \vec n(\vec r,t)=\frac{\vec r-\vec r_0}{|\vec r-\vec r_0|} ~,
 \label{eq:sphervort}
\end{gather}
where $\vec r$ is the spatial part of $x_\nu$ and the profile function $\alpha$ is either one of $\alpha_+, \alpha_-$, which are defined by
\begin{gather}
\alpha_+(r) = \begin{cases} 0 & r < R-\frac{\Delta}{2} \\
     \frac{\pi}{2}\left( 1-\frac{r-R}{\frac{\Delta}{2}} \right) &
     R-\frac{\Delta}{2} < r < R+\frac{\Delta}{2} ~~, \\
                        \pi & R+\frac{\Delta}{2} < r
          \end{cases}\\
\alpha_-(r) = \begin{cases} \pi & r < R-\frac{\Delta}{2} \\
     \frac{\pi}{2}\left( 1+\frac{r-R}{\frac{\Delta}{2}} \right) &
      R-\frac{\Delta}{2} < r < R+\frac{\Delta}{2} ~~. \\
                          0 & R+\frac{\Delta}{2} < r
          \end{cases}
\end{gather}
This means that all links are equal to $\mathbbm 1$ except for the
$t$-links in a single time-slice at fixed $t=1$. The phase changes from 0
to $\pi$ from inside to outside (or vice versa). The graph of $\alpha_-(r)$
for a $40^3 \times N_t$-lattice is plotted in Fig. 2
in~\cite{Jordan:2007ff}, giving a hedgehog-like configuration, without any
intersections and hence no topological charge.  Since only links in the
time direction are different from $\mathbbm 1$, the topological charge
determined from any lattice version of $F \tilde F$ vanishes for this
spherical vortex configuration. We shall refer to such determinations as
topological charge in the rest of this paper.


However, we find a discrepancy to the index of the overlap Dirac
operator~\cite{Narayanan:1994gw,Neuberger:1997fp,Neuberger:1998wv}.
According to the Atiyah-Singer index theorem the topological charge is
related to the index by $\mathrm{ind}\;D[A] = n_- - n_+ = Q$
where  $n_-$ and $n_+$ are the number of left- and right-handed zeromodes
of the Dirac operator \cite{Atiyah:1971rm,Schwarz:1977az,Brown:1977bj}.
For the spherical vortex configuration, Eq.~(\ref{eq:sphervort}), we
obtain a nonzero index.

The lattice version of the index theorem is only valid as long as the gauge
field is smooth enough and satisfies a so-called ``admissibility''
condition. This condition assures that $H^+_L$ has no zero eigenvalues so
that the sign function in the definition of the overlap Dirac operator is
well defined. It requires that the plaquette values $U_{\mu\nu}$ are
bounded close to $\mathbbm 1$, the value for very smooth gauge fields. A
sufficient, but not necessary bound for the ``admissibility'' of the gauge
field is $\mathrm{tr}(\mathbbm 1 - U_{\mu\nu}) < 0.03$ \cite{Luscher:1998du,Neuberger:1999pz,Fukaya:2005ev}

On a  $40^3 \times N_t$-lattice the traces of the plaquettes for the
spherical vortex, Eq.~(\ref{eq:sphervort}), deviate only by a maximum of
1.5\% from trivial plaquettes, satisfying the ``admissibility'' condition.
Nevertheless, the index of the overlap operator is nonzero,
$\mathrm{ind}\;D = \mp 1$, for $\alpha_\pm$.

\begin{figure}
\centering
\includegraphics[scale=.85]{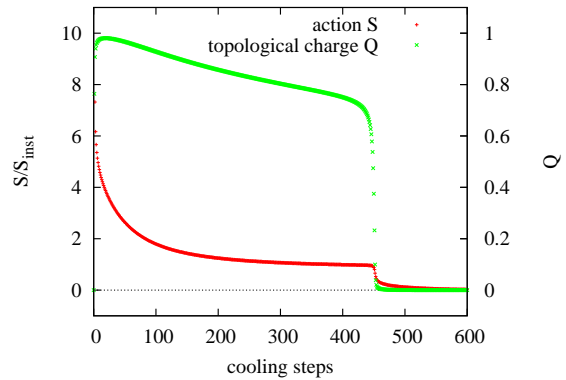}
\caption{Cooling of a spherical vortex on a $40^3 \times 4$ lattice. The
topological charge rises from zero to close to one for $\alpha=\alpha_-$
(right scale) while the action $S$ (in units of the one-instanton action
$S_{\rm inst}$) reaches a plateau (left scale).}
\label{fig:sphcool}
\end{figure}

To try to understand the discrepancy we apply standard cooling to the
spherical vortex configuration. For many cooling steps, the index of the
overlap Dirac operator does not change, but the topological charge quickly
rises close to $\mp 1$ for $\alpha_\pm$  while the action $S$ reaches a
(nonzero) plateau, as shown in Fig.~\ref{fig:sphcool} for a $40^3 \times 4$
lattice. So, the index of the overlap Dirac operator agrees with the
topological charge after some cooling, but not on the original vortex
configuration.

\section{Topology and staggered fermions}
\label{sec:topstagg}

\begin{figure}[h]
\centering
\psfrag{chirality}{abs. chirality ($|\langle\Psi\gamma_5\Psi\rangle|$)}
\psfrag{eigenvalue}{eigenvalue $\lambda$}
\includegraphics[scale=.6]{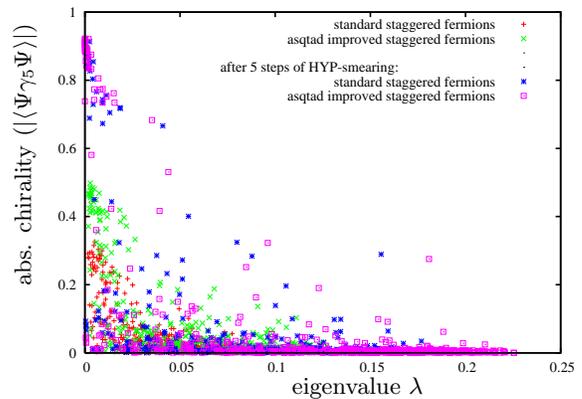}
\caption{The first twenty eigenmodes of thirty Monte Carlo configurations
on a $20^4$ lattice, for standard and asqtad staggered fermions on the
original and HYP-smeared gauge fields, with antiperiodic
boundary conditions.}\label{fig:staggeredzms}
\end{figure}

Staggered fermions don't have exact zeromodes, but a separation between
``would-be'' zeromodes and nonchiral modes is observed for improved
staggered quark actions~\cite{Wong:2004ai}. These results shall be verified
first, by comparing standard and asqtad improved staggered fermions and by
using the HYP (hypercubic blocking) smearing algorithm~\cite{Hasenfratz:2001hp}.
The results are shown in Fig.~\ref{fig:staggeredzms}, presenting the first
twenty eigenmodes of thirty configurations, generated by lattice Monte
Carlo simulation of the tadpole improved L\"uscher-Weisz pure-gauge action
at coupling $\beta_{LW}=3.7$ (lattice spacing $a \approx 0.09$ fm). We
chose a fairly small lattice spacing to keep lattice effects as small as
practicable.  The eigenvalues are plotted against the chirality
(pseudoscalar density) of the modes, which for staggered fermions is
determined by $\langle\Psi\Gamma_5\Psi\rangle$, where $\Gamma_5$
corresponds to a displacement along the diagonal of a hypercube. To ensure
gauge invariance the product includes gauge field multiplications along all
shortest paths connecting opposite corners of the hypercube. Comparing
standard (red $+$ symbols) and asqtad improved (green $\times$ symbols)
staggered fermions, a slight improvement is observed for the latter:
``would-be'' zeromodes show a higher chirality (up to $0.5$) and
eigenvalues are ``closer'' to zero. But for these still rather rough
configurations it seems hard to really identify the would-be zeromodes
reliably. To make the configurations smoother, while retaining topological
fluctuations, we applied five steps of HYP-smearing. Then the chirality of
the would-be zeromodes gets well defined and the eigenvalues get
close to zero. On these smoothed configurations there is no big
difference between standard (blue $*$ symbols) and asqtad improved (magenta
$\Box$ symbols) fermions anymore.

\begin{figure}[t]
\centering
\psfrag{x}{$x$}
\psfrag{y}{$y$}
\psfrag{z}{$z$}
\begin{tabular}{cc}
\includegraphics[width=0.45\columnwidth,keepaspectratio]{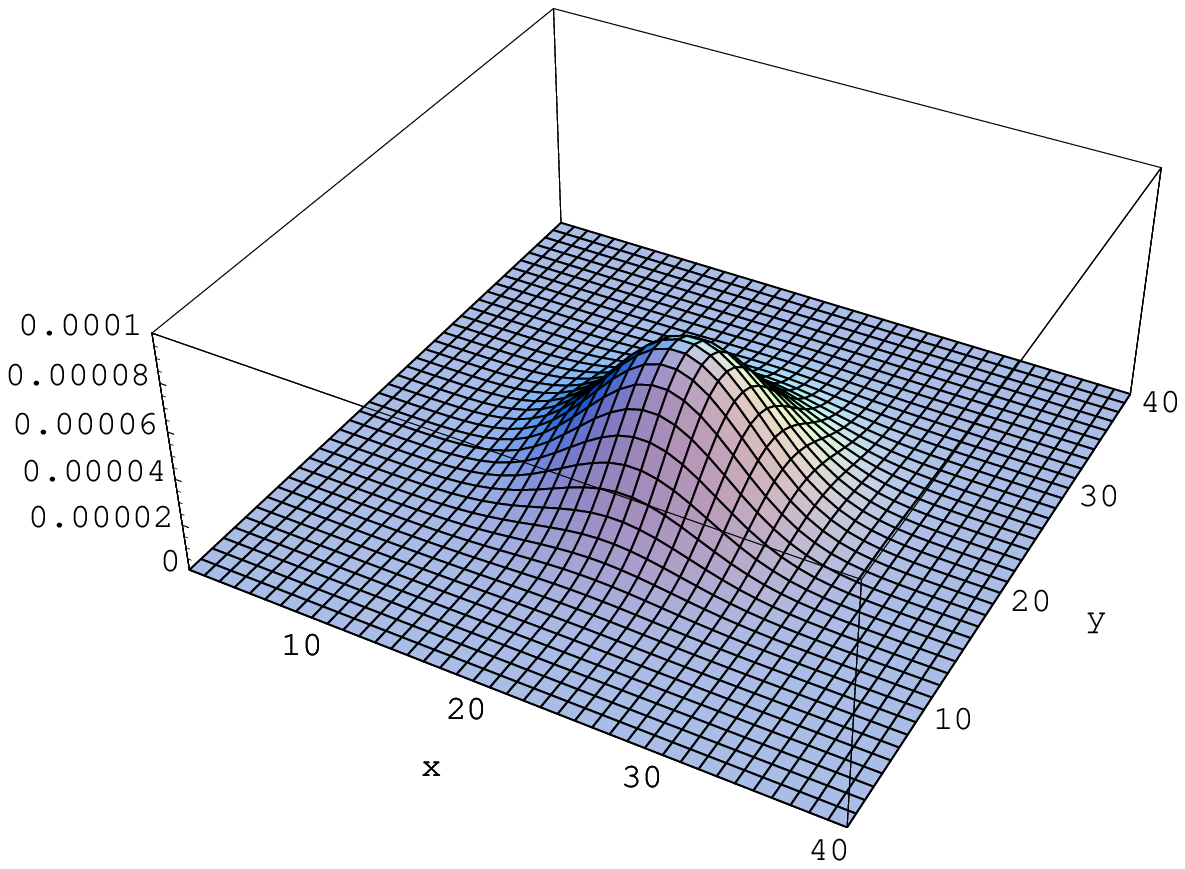} &
\includegraphics[width=0.45\columnwidth,keepaspectratio]{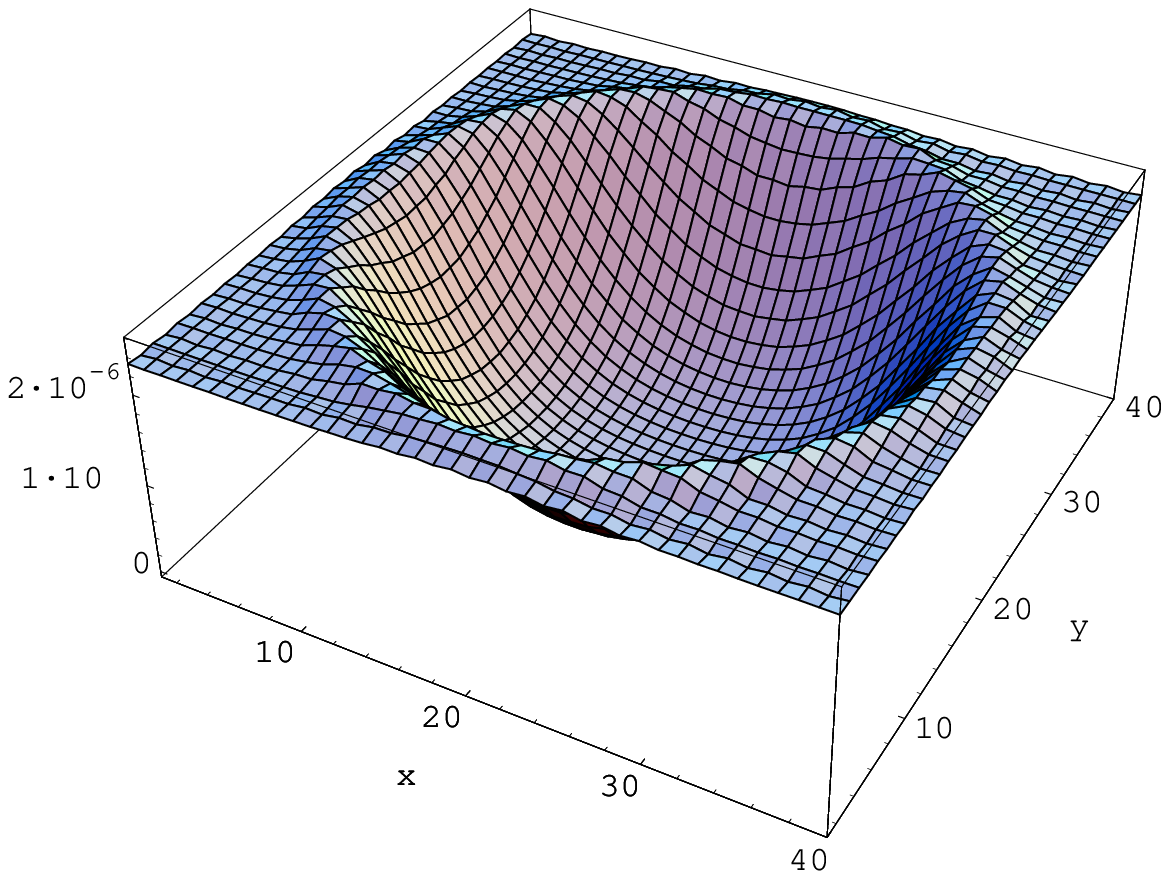}
\end{tabular}
\caption{Scalar densities of asqtad staggered
zeromode of positive chirality for the positive ($\alpha_+$, left) and
negative chirality for the negative ($\alpha_-$, right) spherical vortex in
a time slice of a $40^3 \times 8$ lattice, with antiperiodic boundary conditions.}
\label{fig:stagnegposdens}
\end{figure}

These results promise that one should be able to identify zeromodes
reliably on smooth configurations, such as the above examples of spherical
vortices. The asqtad staggered modes were calculated for a positive and a
negative non-orientable spherical vortex made up of time links in a single
time slice of a $40^3 \times 8$ lattice. Due to charge conjugation, staggered
eigenmodes are doubly degenerated but according to the index theorem
($\mathrm{ind}\;D[A] = n_- - n_+ = 2Q$) we get exactly the same topological charge as the overlap Dirac operator. The scalar densities of
the would-be zeromodes for both types of spherical vortices show a
distribution (Fig.~\ref{fig:stagnegposdens}) that looks very similar to
their overlap counterparts (see~\cite{Jordan:2007ff}).


\section{Index theorem and adjoint fermions}
\label{sec:adjoindex}

It is also interesting to study the correlation of the number of zeromodes
and the topological charge for fermions in the adjoint representation, as
discussed at recent lattice conferences (see {\it e.g.,}
\cite{GarciaPerez:2007ne}). The index of the massless Dirac operator in the
adjoint representation of the $SU(N)$ gauge group in a background field of
topological charge $Q$ is equal to $2NQ$~\cite{Heller:1998cf}. We test the
lattice index theorem for overlap and asqtad staggered fermions in the
adjoint color representation on the spherical vortex configurations, Eq.~(2.1). Since the fermion is in the real representation, the
spectrum of the adjoint Dirac operator is doubly degenerate. Therefore, the
index can only be even valued. It will be interesting whether we find all
possible even values or only multiples of four, resulting in a gauge field
background made up of classical instantons.

In Table~\ref{tab:sphvortzms} the number of positive and negative overlap
and staggered zeromodes in the fundamental and adjoint representation for
periodic and antiperiodic boundary conditions on positive and negative
spherical vortex configurations on different lattice sizes are summarized.
Concerning this table we like to mention two issues. 
All investigated fermion representations equally violate the lattice index
theorem for these special configurations: the index is nonzero while the
topological charge vanishes. 
Further, in some cases, indicated in bold, the adjoint index is not a
multiple of four, as will be discussed in the next sections.

\begin{table}[b]
{\bf negative spherical vortex}\\
\vspace{3mm}
\begin{tabular}{ll}
\quad fundamental: & \qquad\, adjoint:\\
\vspace{1mm}
\quad overlap: \quad\; asqtad: & \qquad\, overlap:  \quad\; asqtad: 
\vspace{1mm}
\end{tabular}\\
\begin{tabular}{rcccccccc}
lattice:&pbc:&apbc:&pbc:&apbc:&pbc:&apbc:&pbc:&apbc:\\ 
$8^4$:&3+4- & 0+1- & 6+8-&  0+2- &  4+\textbf{6-}&  0+\textbf{2-} & 8+\textbf{12-}&  0+\textbf{4-} \\
$12^4$:&3+4-&  0+1- & 6+8-&  0+2- &  4+\textbf{6-}&  0+\textbf{2-}&  8+\textbf{12-}&  0+\textbf{4-} \\
$16^4$:&3+4- & 0+1- & 6+8-&  0+2- &  4+8-&  0+4- & 8+16-&  0+8-\\
$20^4$:&3+4- & 0+1- & 6+8-&  0+2- &  4+8- & 0+4-&  8+16-&  0+8-\\
$40^3\mbox{x}2$:&  3+4- & 0+1- & 6+8- & 0+ 2- &  4+8-&  0+4-&  8+16- & 0+8-\\    
cooled:&3+4-&  0+1-&  6+8- & 0+2- &  4+\textbf{6-} & 0+\textbf{2-}&  8+\textbf{12-}&  0+\textbf{4-}\\
$40^3\mbox{x}4$:&3+4-&  0+1- & 6+8-&  0+2- &  4+8- & 0+4-&  8+16- & 0+8-\\  
\end{tabular}\\
\vspace{8mm}
{\bf positive spherical vortex}\\
\vspace{3mm}
\begin{tabular}{ll}
\quad fundamental: & \qquad\, adjoint:\\
\vspace{1mm}
\quad overlap: \quad\; asqtad: & \qquad\, overlap:  \quad\; asqtad: 
\vspace{1mm}
\end{tabular}\\
\begin{tabular}{rcccccccc}
lattice:&pbc:&apbc:&pbc:&apbc:&pbc:&apbc:&pbc:&apbc:\\ 
$8^4$:&1+0-&  4+3-&  2+0-&  8+6- &  \textbf{6+}4- & \textbf{2+}0- & \textbf{12+}8- & \textbf{4+}0-\\
$12^4$:&1+0-&  4+3-&  2+0-&  8+6- &  \textbf{6+}4- & \textbf{2+}0- & \textbf{12+}8-& \textbf{4+}0-\\
$16^4$:&1+0-&  4+3-&  2+0-&  8+6- &  8+4- & 4+0- & 16+8- & 8+0-\\
$20^4$:&1+0-&  4+3-&  2+0-&  8+6-&   8+4- & 4+0- & 16+8-&  8+0-\\
$40^3\mbox{x}2$:&1+0-&  4+3- & 2+0- & 8+6- &  8+4- & 4+0-&  16+8- & 8+0-\\
   cooled: & 1+0-&  4+3- & 2+0- & 8+6- &  \textbf{6+}4- & \textbf{2+}0-& \textbf{12+}8- & \textbf{4+}0-\\
$40^3\mbox{x}4$:&1+0-&  4+3-&  2+0- & 8+6- &  8+4- & 4+0- & 16+8- & 8+0-\\
\vspace{1mm}
\end{tabular}
\caption{Number of positive and negative overlap and staggered zeromodes in
the fundamental and adjoint representation for periodic and antiperiodic
boundary conditions on positive and negative spherical vortices on
different lattice sizes. Numbers which indicate half-integer topological
charge are printed in bold.}
\label{tab:sphvortzms}
\end{table}

To study the above mentioned violation, we consider the homotopy of the above spherical vortex configurations, Eq.~(2.1). The $t$-links of these spherical vortices fix the holonomy of the time-like loops, defining a map $U_t(\vec x,t=1)$ from the $xyz$-hyperplane at $t=1$ to $SU(2)$. Because of the periodic boundary conditions, the time-slice has the topology of a 3-torus. But, actually, we can identify all points in the exterior of the 3 dimensional sphere since the links there are trivial. Thus the topology of the time-slice is $\mathbbm R^3 \cup \{\infty\}$ which is homeomorphic to $S^3$. A map $S^3 \to SU(2)$ is characterised by a winding number 
\begin{gather}
  \cal N = -\frac{1}{24\pi^2} \int \mathrm{d}^3 x \,\epsilon_{ijk}\,\mbox{tr}[(U^\dag\partial_i U)(U^\dag\partial_j U)(U^\dag\partial_k U)],
\end{gather}
resulting in $\cal N=-1$ for positive and $\cal N=+1$ for negative spherical vortices.
With this assignment the index of the Dirac operator and the topological charge after cooling coincide with this winding number $\cal N$. Obviously such windings, given by the holonomy of the time-like loops of the spherical vortex, influence the index theorem~\cite{Nye:2000eg,Poppitz:2008hr}.

\clearpage

\section{Adjoint zeromodes and fractional topological charge}
\label{sec:fractadjoint}

Classical instantons carry an integer topological charge. Thus, in case of
a fermion in the fundamental representation of $SU(2)$ there is exactly one
zeromode for a one-instanton configuration. Now, if the actual constituents
of the QCD vacuum had topological charge $Q=1/2$, no zeromode would be
produced. However, for adjoint fermions configurations with topological
charge $Q=1/2$ are able to create a zeromode. Edwards et al.\ presented
in~\cite{Heller:1998cf} some evidence for fractional topological charge on
the lattice. Garc\'ia-P\'erez et al.~\cite{GarciaPerez:2007ne}, however,
associated this to lattice artefacts, {\it i.e.,} topological objects of
size of the order of the lattice spacing. For comparison we repeat the
calculations with Monte Carlo configurations on $12^4$ lattices for overlap
and $16^4$ lattices for asqtad staggered fermions. In the first case we
find for five out of twenty configurations a number of overlap zeromodes
which is not a multiple of four, {\it i.e.,} 2, 6, 6, 10, 10. For adjoint
asqtad staggered fermions multiples of eight are expected, since staggered
fermions have a doubly degenerate eigenvalue spectrum.  Nevertheless we
even find a couple of configurations with one or two ``zeromodes'', but of
course it is not clear whether all these low-lying modes are would-be
zeromodes or whether all would-be zeromodes were identified.

However, the measurements with adjoint fermions on the classical spherical
vortices also show fractional topological charge for small lattice sizes,
see Table~\ref{tab:sphvortzms}. Very interesting is the case of the cooled
$40^3 \times 2$ lattice configuration, which will be discussed next.

\section{Cooling of vortex configurations}
\label{sec:fracttconfig}

\begin{figure}[h]
\centering
\includegraphics[scale=1.]{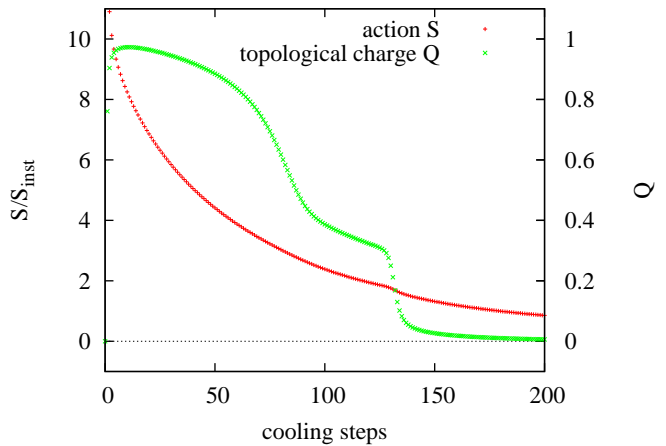}
\caption{Cooling of a spherical vortex with $\alpha=\alpha_-$ on a $40^3
\times 2$ lattice. While the action $S$ (in units of the one-instanton
action $S_{\rm inst}$) decreases slowly (left scale), the topological
charge first rises from zero to close to one (right scale), then decreases
to an intermediate plateau of $Q\approx0.4$ before it vanishes.}
\label{fig:sphcool2}
\end{figure}

\begin{figure}
\centering
\begin{tabular}{ccc}
\includegraphics[scale=.45]{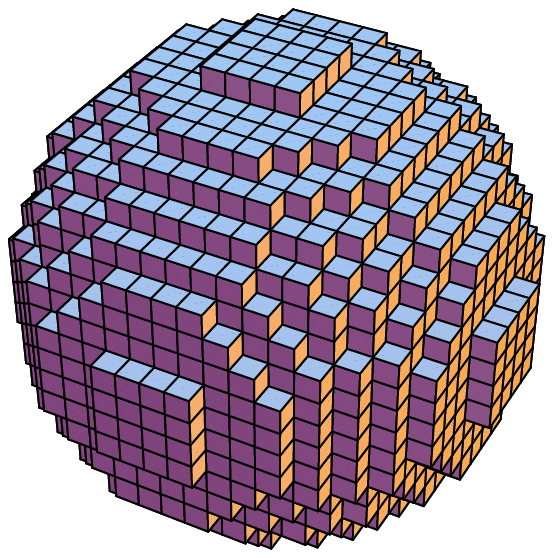}&\includegraphics[scale=.35]{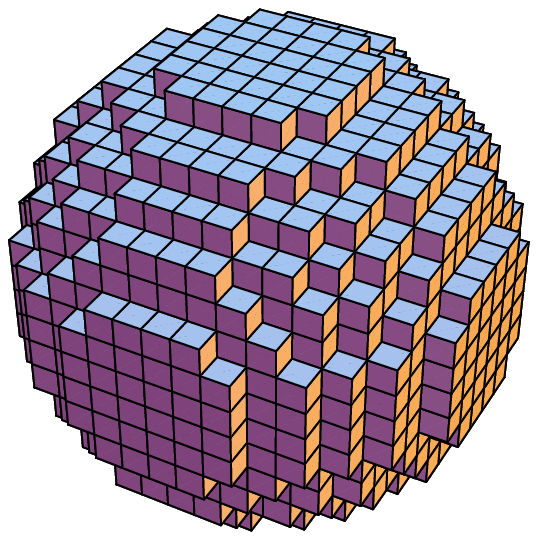}&\includegraphics[scale=.26]{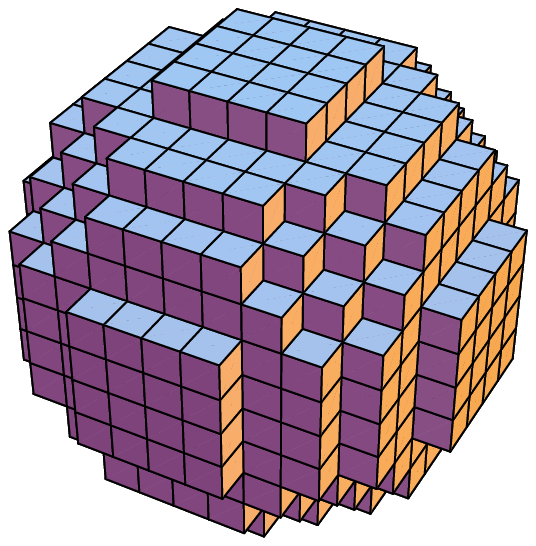}\\
0 cooling steps&20 cooling steps&40 cooling steps\\
\includegraphics[scale=.14]{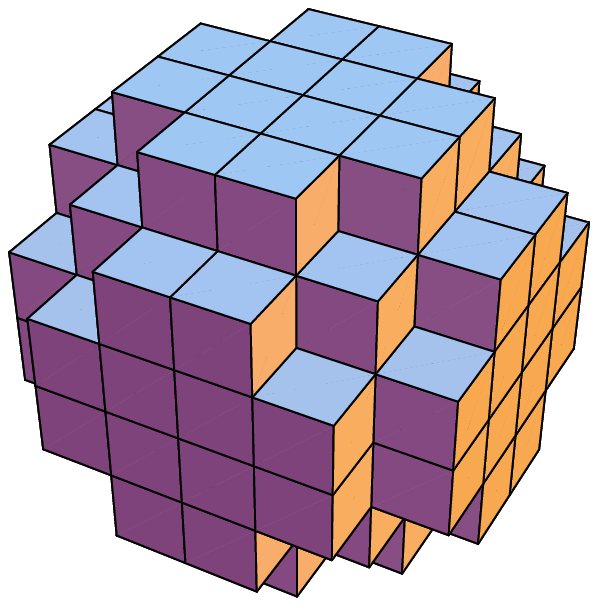}&\includegraphics[scale=.1]{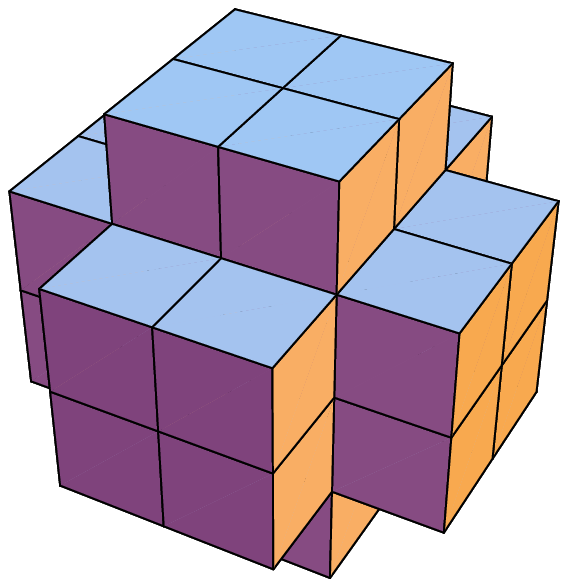}&\includegraphics[scale=.05]{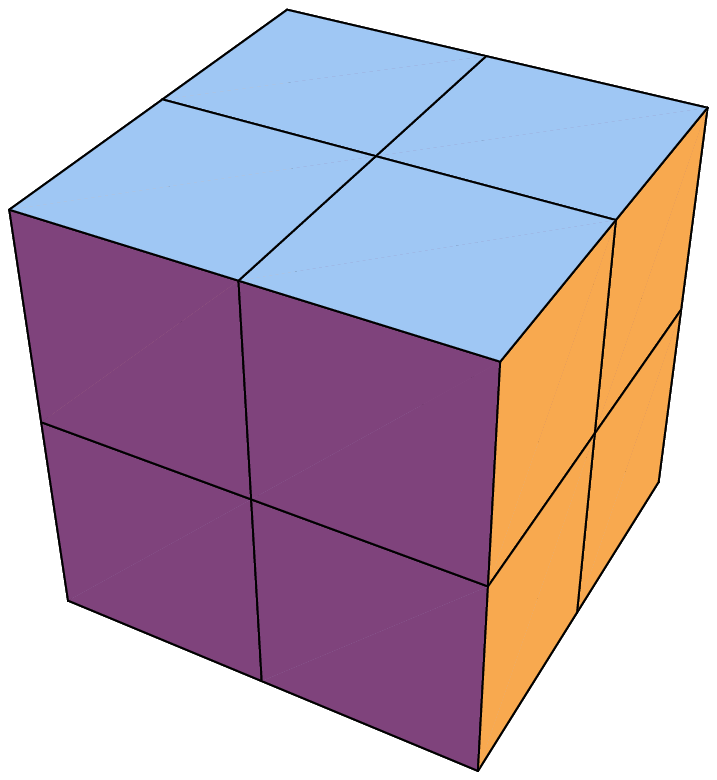}\\
60 cooling steps&70 cooling steps&78 cooling steps\\
\end{tabular}
\caption{Cooling of a spherical vortex on a $40^3
\times 2$ lattices. The vortex shrinks and vanishes after 78 cooling
steps.}
\label{fig:vortcontract}
\end{figure}

\begin{figure}[t]
\centering
\includegraphics[scale=.59]{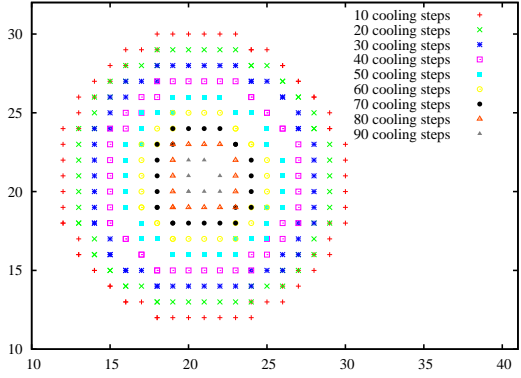}
\caption{Cooling of a spherical vortex with $\alpha=\alpha_-$ on a $40^3
\times 2$ lattice. In maximal abelian gauge a monopole-antimonopole circle
shrinks and vanishes at 92 cooling steps.}
\label{fig:moncontract}
\end{figure}

During cooling of a spherical vortex on a $40^3 \times 2$ lattice we find
evidence for fractional topological charge as shown in
Fig.~\ref{fig:sphcool2}. It exhibits a second ``plateau'' for the
topological charge with $Q \approx 1/2$ between cooling steps 100 and 130.
We analyzed this configuration with adjoint eigenmodes, which indeed
measure fractional topological charge, as seen in
Table~\ref{tab:sphvortzms}. The spherical vortex contracts during cooling
as shown in Fig.~\ref{fig:vortcontract}. After 78 cooling steps the vortex
structure vanishes. In the maximal abelian gauge one can identify a
monopole-antimonopole ring after projection which again contracts during
cooling and disappears after 91 cooling steps, see
Fig.~\ref{fig:moncontract}. In the interesting region with topological
charge $Q=1/2$ (at cooling step 120), the action and topological charge
densities concentrate in the center of the spacial volumes and the gauge
field at the lattice boundary is trivial.  Landau gauge yields a very
symmetric configuration, a static, singular monopole without an
antimonopole.
%
%
In Landau gauge both time slices are exactly the same and all time-like
plaquettes are trivial. The central cube in every time-slice is a
non-abelian representation~\cite{Wu:1975vq,Faber:2002nw} of a Dirac
monopole, see also Fig. 9. Its six plaquettes correspond to rotations by $2\pi/3$ in the
fundamental representation of $SU(2)$. In the color frame of the corner
with the smallest coordinate values the plaquette color vectors,
parallel-transported according to Fig. 2 in~\cite{Skala:1996ar}, point in
the $(-1,-1,-1)$-direction. Hence, the six plaquettes sum up to a total
rotation of $4\pi$ in agreement with the non-abelian lattice Bianchi
identity, which states that the product of plaquette rotations in
appropriate order results in the unit matrix. Nevertheless the magnetic
flux out of the cube is nonvanishing~\cite{Skala:1996ar}.

In the $U(1)$ representation in color direction $(1,1,1)$ the central cube
represents a Dirac monopole with plaquette values summing up to $2\pi$. The
link variables of the central cube correspond to an $SU(2)$ rotation of a
unit vector from one corner to the other, {\it i.e.,}
$\cos(\omega)=1/3=(-1,-1,-1)(1,-1,-1)/3$. Due to the non-abelian nature of
the $SU(2)$ gauge field for increasing distance from the center the field
strength approaches zero. No antimonopole is needed to compensate the
monopole.

\begin{figure}[h]
\centering
\psfrag{Polyakov}{Polyakov}
\psfrag{cooling steps}{cooling steps}
\psfrag{(x,20,20)}{(x,20,20)}
\includegraphics[width=.9\linewidth]{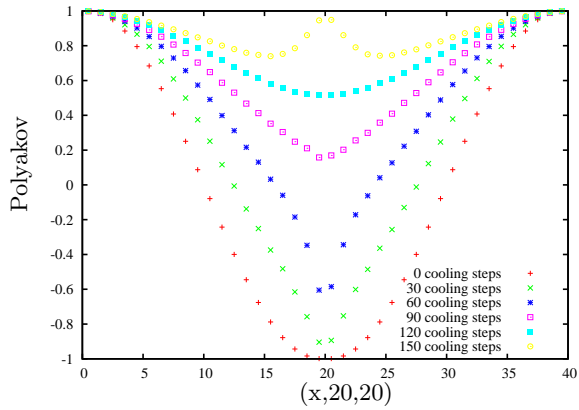}
\caption{Profile functions of the Polyakov loops through the lattice
(varying $x$ at $y=20$ and $z=20$) during cooling.}
\label{fig:polprofile}
\end{figure}
\begin{figure}[h]
\centering
\psfrag{fig}{FIG. 9: The arrows indicate the rotational vectors of link-,
plaquette- and Polyakov-matrices around the central cube of the}
\psfrag{fit}{negative spherical vortex after 120 cooling steps. Link-vectors rotate around
the center, color fluxes through plaquettes are}
\psfrag{fil}{aligned parallel in the $(-1,-1,-1)$-direction and Polyakov loops form a hedgehog.}
\includegraphics[width=2\linewidth]{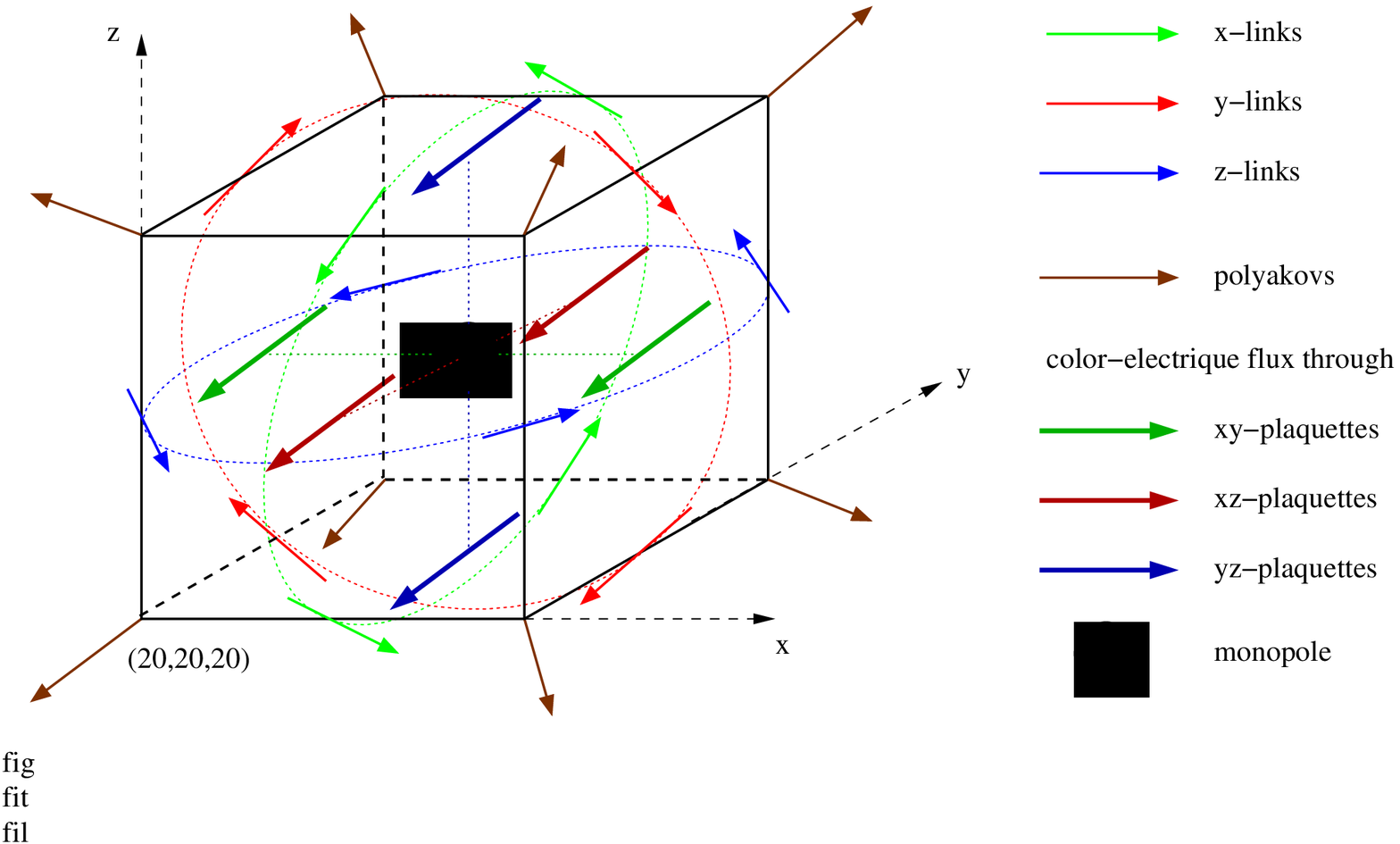}
\end{figure}
\begin{figure}[h]
\centering
\psfrag{Polyakov}{Polyakov}
\psfrag{cooling steps}{cooling steps}
\psfrag{(x,20,20)}{(x,20,20)}
\includegraphics[width=.9\linewidth]{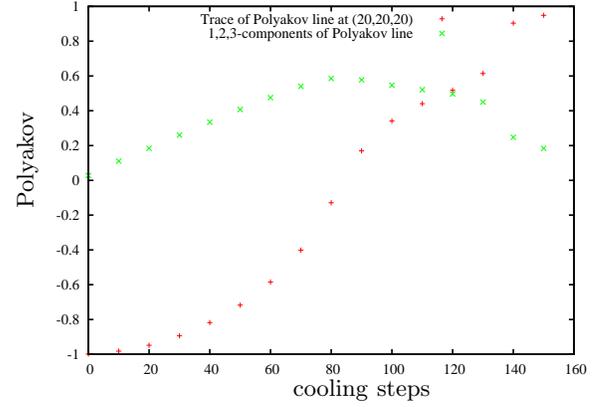}
\caption{Polyakov-line at spacial site (20,20,20) for the negative
spherical vortex during cooling shows two transitions as function of
the cooling step.}
\label{fig:pollauf}
\end{figure}    
\begin{table}[h]
\begin{tabular}{p{\linewidth}}
\normalsize
The Polyakov loops around the central cube form a hedgehog. They reflect
the color directions of the link variables. A parallel-transport to the
above color frame leads to parallel Polyakov lines in color direction
$(-1,-1,-1)$. The profile function of the Polyakov loops through the
lattice during cooling is shown in Fig.~\ref{fig:polprofile} and the course
of the central Polyakov line is plotted in Fig.~\ref{fig:pollauf}. We find
two transitions, one smooth between cooling steps 70 and 90 and a jump
after 130 cooling steps, where the monopole finally vanishes. The color
directions of link-, plaquette- and Polyakov matrices of the central 
cube of the negative spherical vortex after 120 cooling steps are depicted in Fig. 9.
\vspace{10cm}

\phantom{.}
\end{tabular}
\end{table}

\clearpage

\section{Conclusions}
\label{sec:conclusio}

We reported on violations of the lattice index theorem for smooth,
``admissible'' gauge configurations of classical, spherical center
vortices, for both, overlap and asqtad staggered fermions in the
fundamental and adjoint representations. Numerically, the discrepancy 
equals the winding number of the spheres when they are regarded as 
maps $S^3 \to SU(2)$. 

During cooling a spherical vortex on a $40^3 \times 2$ lattice, 
the index in the adjoint representation indicates evidence for an 
object with fractional topological charge, which is identified as a 
Dirac monopole with the well known singularity in its center and a 
gauge field fading away at large distances. Therefore even for periodic 
boundary conditions it does not need an antimonopole.

\acknowledgments{We thank Rob Pisarski and Jeff Greensite for the
suggestion to investigate configurations with fractional topological charge
with adjoint fermions. We are grateful to Mithat \"Unsal and {\v S}tefan Olejn\'{\i}k for helpful discussions. This research was partially supported by the Austrian
Science Fund (``Fonds zur F\"orderung der Wissenschaften'', FWF) under
contract P22270-N16 (R.H.).}

\bibliographystyle{unsrt}
\bibliography{../literatur}

\end{document}